\documentclass[12pt,,letterpaper]{JHEP3}
\usepackage{graphics}
\usepackage{epsfig}
\usepackage{amssymb}
\usepackage{stmaryrd}
\usepackage{amsmath}
\usepackage{amsfonts}
\usepackage{mathrsfs}
\usepackage{graphicx}

\def\be{\begin{equation}}
\def\ee{\end{equation}}
\def\ba{\begin{eqnarray}}
\def\ea{\end{eqnarray}}

\title{Poor man's holography: How far can it go?}
\author{Yu Tian\\
College of Physical Sciences, Graduate University of Chinese Academy
of Sciences, Beijing 100049, China\\
\email{ytian@gucas.ac.cn}}

\author{Xiao-Ning Wu\\
Institute of Mathematics, Academy of Mathematics and System Science,
CAS, Beijing 100190, China\\
\email{wuxn@amss.ac.cn}}

\author{Hongbao Zhang\\
Crete Center for Theoretical Physics, Department of Physics, \\
University of Crete, 71003 Heraklion, Greece\\
 \email{hzhang@physics.uoc.gr}}

\preprint{CCTP-2012-10}

\abstract{Almost a century ago Einstein, after Newton, shed a new light on gravity by claiming that gravity is geometry. There has been no deeper insight beyond that later on except the recent suspicion that gravity may also be holographic, dual to some sort of quantum field theory living on the boundary with one less dimension. Such a suspicion has been supported mainly by a variety of specific examples from string theory. This paper is intended to purport the holographic gravity from a different perspective. Namely we shall show such a holography can actually be observed  by working merely within the context of Einstein's gravity through promoting Brown-York's formalism, where neither is the spacetime required to be asymptotically AdS nor the boundary to be located at conformal infinity, which also conforms to the spirit inherited from Wilson's effective field theory. In particular, we show that our holography works remarkably well at least at the level of thermodynamics and hydrodynamics, where a perfect matching between the bulk gravity and boundary fluid is found for entropy and its production by the conserved current method.}

\begin{document}

\section{Introduction}

Although string theory provides an explicit implementation of quantum gravity in a holographic way, now dubbed as AdS/CFT correspondence, it is worthwhile to keep in mind that there are various hints from within the context of Einstein's gravity towards the speculation that gravity is essentially holographic, where neither supersymmetry nor strings as well as branes are involved. Here we would like list four of them as follows.

Brown-Henneaux's asymptotic symmetry analysis for three dimensional gravity\cite{BH}.

Brown-York's surface tensor formulation of quasilocal energy and conserved charges\cite{BY}.

Black hole thermodynamics\cite{Wald1}.

Bousso's covariant entropy bound\cite{Bousso}.

In particular, Brown-York's surface tensor formulation bears a strong resemblance to the recipe in the dictionary for AdS/CFT correspondence especially when one is brave enough to declare that Brown-York's surface tensor is not only for the purpose of the bulk side but also for some sort of system living on the boundary.  In this sense, Brown-York's tensor formulation implies that gravity is holographic. Actually, such a holographic interpretation can be exactly proven at least at the level of thermodynamics and hydrodynamics. This is the main purpose of this paper.

Let us first promote such a formulation in a holographic way by the following dictionary, i.e.,
\begin{equation}
\int_{\phi_0} D\phi\exp(-S_{bulk}[\phi])=\int D\psi\exp(-I_{FT}[\phi_0,\psi]),
\end{equation}
where $\phi_0$ plays a dual role, namely, serves as the boundary condition for the bulk path integral over $\phi$  on the left handed side and as the external background of dual boundary field theory on the right handed side. When the spacetime is asymptotically AdS with the conformal boundary, the above dictionary recovers the standard AdS/CFT correspondence. But here we do not require the spacetime to be asymptotically AdS, or  the boundary to be located at the conformal infinity. Furthermore, in the saddle point approximation, the expectation value of dual operator is given by
\begin{equation}
\langle O\rangle\equiv\frac{1}{\sqrt{-\gamma}}\frac{\delta I_{FT}[\phi_0,\psi]}{\delta\phi_0}=\frac{1}{\sqrt{-\gamma}}\frac{\delta S_{classical}}{\delta\phi_0}.
\end{equation}
Two examples are of special interest. One is the case of $\phi$ to be the bulk metric $g_{ab}$ with $\phi_0$ the induced metric $\gamma_{ab}$ on the boundary, in which the dual operator is simply the stress-energy tensor of the boundary system, and its expectation value is given by
\begin{equation}
t^{ab}=\frac{2}{\sqrt{-\gamma}}\frac{\delta S_{classical}}{\delta\gamma_{ab}}.
\end{equation}
The other is the case of $\phi$ to be the electromagnetic potential $A_a$ with $\phi_0$ the pull back of $A_a$ on the boundary, in which the dual operator is just the electric current with its expectation value given by
\begin{equation}
j^a=\frac{1}{\sqrt{-\gamma}}\frac{\delta S_{classical}}{\delta A_a}.
\end{equation}
In particular, if the bulk action for gravity and electromagnetic fields are given by Einstein-Hilbert action plus Gibbons-Hawking term and Maxwell action respectively, i.e.,
\begin{eqnarray}
S_{GR}&=&\frac{1}{16\pi}[\int d^{d+1}x\sqrt{-g}(R-2\Lambda)+2\int d^dx\sqrt{-\gamma}K], \nonumber\\
S_{EM}&=&-\frac{1}{16\pi}\int d^{d+1}x\sqrt{-g}F_{ab}F^{ab},
\end{eqnarray}
we have
\begin{equation}
t^{ab}=\frac{1}{8\pi}(K\gamma^{ab}-K^{ab}-C\gamma^{ab}), j^a=-\frac{1}{4\pi}n_bF^{ba},
\end{equation}
where $K=\gamma^{ab}K_{ab}$ is the trace of extrinsic curvature $K_{ab}=\gamma_a^c\nabla_cn_b$ with $n_b$ the outward normal vector to the boundary, and $C$ the constant from some sort of renormalization.

\section{Equilibruim state version of correspondence: thermodynamics}

First of all, let us build up the equilibrium state version of our correspondence by considering the Schwarzschild AdS black hole in the bulk, i.e.,
\begin{equation}
ds^2_{d+1}=\frac{dr^2}{f(r)}-f(r)dt^2+r^2d\Omega^2_\varepsilon, f(r)=\varepsilon+\frac{r^2}{L^2}-\frac{2M}{r^{d-2}},
\end{equation}
where $d\Omega^2_\varepsilon$ can be the metric on the sphere, plane or hyperbola for $\varepsilon=1,0,-1$ respectively. Then by the standard calculation, the entropy and temperature of black hole are given by
\begin{equation}
S_{BH}=\frac{r_h^{d-1}\Omega_\varepsilon}{4}, T_H=\frac{f'(r_h)}{4\pi}
\end{equation}
with $r_h$ the location of horizon satisfying $f(r_h)=0$.

Now the boundary can be any hypersurface of $r=r_c$ outside the horizon, with the induced metric 
\begin{equation}
ds^2_d=-f_cdt^2+r_c^2d\Omega_\varepsilon^2, f_c=f(r_c).
\end{equation}
The nice thing is that one can easily show that the boundary stress-energy tensor has a form of ideal fluid, i.e.,
\begin{equation}
t^{ab}=\epsilon u^au^b+p(u^au^b+\gamma^{ab})
\end{equation}
with the fluid four velocity $u^a=\frac{1}{\sqrt{f_c}}(\frac{\partial}{\partial t})^a$ on the boundary,  and the energy density as well as pressure given by
\begin{eqnarray}
\epsilon&=&-\frac{(d-1)\sqrt{f_c}}{8\pi r_c}+C,\nonumber\\
p&=&\frac{(d-2)\sqrt{f_c}}{8\pi r_c}+\frac{f'_c}{16\pi \sqrt{f_c}}-C.
\end{eqnarray}
Note that the volume of the boundary system is 
\begin{equation}
V=r_c^{d-1}\Omega_\varepsilon.
\end{equation}
So the total energy is given by
\begin{equation}
E=\epsilon V=(-\frac{(d-1)\sqrt{f_c}r_c^{d-2}}{8\pi}+Cr_c^{d-1})\Omega_\varepsilon.
\end{equation}
If our bulk/boundary correspondence is right, we must have the black hole entropy identified as the entropy for the boundary fluid, i.e.,
\begin{equation}
S_{BF}=S_{BH}.
\end{equation}
With this identification, we can express $E$ as a function of $S_{BF}$ and $V$, which further gives rise to
\begin{equation}
\frac{\partial E}{\partial S_{BF}}=T_c, \frac{\partial E}{\partial V}=-p,
\end{equation}
where $T_c=\frac{T_H}{\sqrt{f_c}}$ is the temperature for the boundary fluid, redshifted as it should be the case. So we have a well defined first law of thermodynamics for the boundary fluid, i.e.,
\begin{equation}
dE=T_cdS_{BF}-pdV.
\end{equation}

\section{Physical process version of correspondence: hydrodynamics}

If the boundary system is perturbed by some sort of external sources away from the equilibrium state, then the transport process will intend to bring the system back to a new equilibrium state, which generically causes entropy production. In particular, when the boundary system is perturbed by the electromagnetic field $A_an^a=0$ and gravitational field satisfying $h_{ab}n^a=0$ as well as $h_{00}=0$, the rate for entropy production is given by\footnote{With such a setup, actually the second term consists of two contributions, namely  the entropy production induced by the inhomogeneous temperature\cite{Hartnoll,Herzog,TWZ}, and the one produced by the shear as well as bulk viscosity.}\cite{Reichl}
\begin{equation}
\Sigma=\frac{1}{T_c}j^cE_c-\frac{1}{T_c}t^{(1)ab}(D_a^{(1)}u_b+D_au_b^{(1)}).
\end{equation}
Here the superscript $1$ denotes the first order variation induced by the gravitational perturbation $h_{ab}$. For instance,
\begin{equation}
D_a^{(1)}u_b+D_au_b^{(1)}=-\Gamma^{(1)c}{}_{ab}u_c+D_a(h_{bc}u^c)=\sqrt{f_c}(\Gamma^{(1)0}{}_{bc}-D_ah_b^0),
\end{equation}
where we have used the fact that $D_a^{(1)}$ comes essentially from  the first order variation of Christoffel symbol, i.e.,
\begin{equation}
D_a^{(1)}v^b=\Gamma^{(1)b}{}_{ac}v^c=\frac{1}{2}\gamma^{bd}(D_ah_{cd}+D_ch_{ad}-D_dh_{ac})v^c.
\end{equation}

From the bulk point of view, such perturbations on the boundary
should propagate towards the black hole and be absorbed. Eventually the black hole will settle down to a new static final state with an increase
in the area of the black hole horizon, or put it another way, with an increase of black hole entropy. If our bulk/boundary correspondence is right, the increase of black hole entropy should be precisely equal to the aforementioned entropy production on the boundary. As we shall prove shortly, this is actually the case. The basic idea for such a proof is to relate the boundary to the bulk by the conserved current, which can be best presented by considering first the electromagnetic perturbation.

Let us start with the stress-energy tensor for the electromagnetic waves, i.e.,
\begin{equation}
T^{ab}_{EM}=\frac{2}{\sqrt{-g}}\frac{\delta S_{EM}}{\delta g_{ab}}=\frac{1}{4\pi}(F^{ac}F^b{}_c-\frac{1}{4}g^{ab}F_{cd}F^{cd}),
\end{equation}
which is conserved $\nabla_aT^{ab}_{EM}=0$. So one can construct the conserved current as follows
\begin{equation}
J^a=T^{ab}_{EM}\xi_b
\end{equation}
associated with the timelike Killing vector field $\xi=\frac{\partial}{\partial t}$. Now suppose the
non-equilibrium region has compact support on the boundary, which naturally gives rise to the corresponding compact support for both of the perturbed bulk and perturbed horizon. Then  integrating $\nabla_aJ^a=0$ over the perturbed bulk with the perturbed horizon as the inner boundary and using Gauss law, we end up with
\begin{equation}
\int_HT^{ab}_{EM}\xi_a\xi_b=\int_\mathrm{bdry}T^{ab}_{EM}n_a\xi_b,
\end{equation}
where $H$ is the horizon. To relate the left handed side with the increase in the black hole entropy in a simple way, we would like to make the null geodesic generators of the event horizon of the perturbed black hole coincide with the null geodesic generators of the unperturbed black hole by using our diffeomorphism freedom\cite{GW}. With this, the perturbation in the horizon location vanishes and $\delta\xi\propto\xi$ on the horizon. Then Raychaudhuri equation implies\cite{Wald2,Poisson}
\begin{equation}
T_H\delta S_{BH}=\int_HT^{ab}_{EM}\xi_a\xi_b.
\end{equation}
On the other hand, with the electric field felt by the boundary fluid as $E_c=F_{cb}u^b$, we have
\begin{equation}
\int_\mathrm{bdry}T^{ab}_{EM}n_a\xi_b=\sqrt{f_c}j^aE_a,
\end{equation}
which gives rise to\footnote{Obviously, in order to guarantee the increase of entropy, one is forced to impose the natural boundary condition for the perturbation on our cutoff surface by requiring the conserved current flux be allowed only from the outside to the interior bulk. }
\begin{equation}
\delta S_{BH}=\frac{j^aE_a}{T_c}=\delta S_{BF}.
\end{equation}

Next we consider the case for the entropy production induced by the gravitational perturbation on the boundary. To proceed, let us first expand the bulk Einstein equation
on the black hole background to second order, i.e.,
\begin{eqnarray}
G^{ab}+\Lambda g^{ab}&=&0,\\
G^{(1)ab}[h]-\Lambda h^{ab}&=&0,\label{dual1}\\
G^{(1)ab}[q]-\Lambda q^{ab}&=&8\pi T^{ab}_{GW}[h]=-[G^{(2)ab}[h]+\Lambda h^a{}_ch^{cb}],\label{dual2}
\end{eqnarray}
where the metric is expanded as $g_{ab}+\epsilon h_{ab}+\epsilon^2 q_{ab}$ with the indices raised or lowered by the background metric $g_{ab}$. Furthermore, it follows from Bianchi identity that the energy-momentum tensor is conserved for the gravitational waves propagating on the background, i.e.,
\begin{eqnarray}
\nabla_aT^{ab}_{GW}=0,
\end{eqnarray}
which, as before, gives rise to
\begin{equation}
\delta S_{BH}=\frac{1}{T_H}\int_H T^{ab}_{GW}\xi_a\xi_b=\frac{1}{T_H}\int_\mathrm{bdry}T^{ab}_{GW}n_a\xi_b.
\end{equation}
So now the task boils down into whether one can express the above flux across the boundary in terms of entropy production on the boundary, which can actually be achieved by a straightforward but lengthy calculation. But here we would like to present a shortcut towards the final result by taking advantage of the dual role played by the gravitational waves. Namely, as demonstrated in Eqs.(\ref{dual1}) and (\ref{dual2}), the gravitational waves, albeit treated as sort of matter waves like light, are essentially ripples in the fabric of spacetime. Thus we can relate the aforementioned flux to the quantities for the dual system on the boundary by Gauss-Codazzi equation, which, expanded to second order, gives
\begin{eqnarray}
D_at^{ac}&=&-\frac{1}{8\pi}G^{ab}n_a\gamma_b^c=0,\\
D_at^{(1)ac}+D_a^{(1)}t^{ac}&=&-\frac{1}{8\pi}G^{(1)ab}[h]n_a\gamma_b^c=0,\\
D_at^{(2)ac}+D_a^{(1)}t^{(1)ac}+D_a^{(2)}t^{ac}&=&-\frac{1}{8\pi}G^{(2)ab}[h]n_a\gamma_b^c=T^{ab}_{GW}n_a\gamma_b^c,
\end{eqnarray}
where $D_a^{(2)}$ is determined by the second order Christoffel symbol, i.e.,
\begin{equation}
D_a^{(2)}v^b=\Gamma^{(2)b}{}_{ac}v^c=-\frac{1}{2}h^{bd}(D_ah_{cd}+D_ch_{ad}-D_dh_{ac})v^c.
\end{equation}
Then one can show
\begin{eqnarray}
\delta S_{BH}&=&\frac{1}{T_H}\int_\mathrm{bdry}T^{ab}_{GW}n_a\xi_b=-\frac{\sqrt{f_c}}{T_c}\int_\mathrm{bdry}\frac{1}{2}D_dht^{(1)d0}+\Gamma^{(1)0}{}_{cd}t^{(1)cd}+\Gamma^{(2)0}{}_{cd}t^{cd}\nonumber\\
&=&-\frac{\sqrt{f_c}}{T_c}\int_\mathrm{bdry}\frac{1}{2}hD_d^{(1)}t^{d0}+\Gamma^{(1)0}{}_{cd}t^{(1)cd}+\Gamma^{(2)0}{}_{cd}t^{cd}\nonumber\\
&=&-\frac{\sqrt{f_c}}{T_c}\int_\mathrm{bdry}-\frac{1}{4}h\gamma^{c0}D_ch_{ad}t^{ad}-\frac{1}{2}h_a^0D_dht^{ad}+\Gamma^{(1)0}{}_{cd}t^{(1)cd}+\Gamma^{(2)0}{}_{cd}t^{cd}\nonumber\\
&=&-\frac{\sqrt{f_c}}{T_c}\int_\mathrm{bdry}-\frac{1}{2}h_a^0D_dht^{ad}+\Gamma^{(1)0}{}_{cd}t^{(1)cd}+\Gamma^{(2)0}{}_{cd}t^{cd}
=\delta S_{BF}.
\end{eqnarray}
where we have thrown away all the total derivative terms at each step, and employed $D_ct^{ad}=0$ as well as $h_{ad}t^{ad}=ph$ in the second last step.

\section{Discussion}
We have provided Brown-York's  formalism with the holographic interpretation. In particular, we have demonstrated that such a holographic formulation works very well at least at the level of thermodynamics and hydrodynamics, where a perfect matching between the bulk gravity and boundary system is exactly derived for entropy and its production. Although we are working only with Schwarzschild AdS black hole, it can be shown that our discussion can be applied to charged AdS black hole, where the calculation is a little bit involved due to the fact that the electromagnetic and gravitational perturbations are coupled to each other\cite{TWZ}. Furthermore,  it is obvious that our procedure can actually be applied not only to asymptotically flat and De-Sitter charged black holes but also to the spacetime patch associated with Rindler horizon in the flat spacetime or De-Sitter horizon in the De-Sitter spacetime\cite{TWZ}.

Compared to the standard AdS/CFT correspondence,  our holography is more general in the following sense. First, we do not require the spacetime to be asymptotically AdS. Second, our boundary is not required to be located at conformal infinity. Actually it can be found an echo for such a relaxation in the membrane paradigm\cite{Damour,PT,KSS,IL,BKLS}. In addition, such a generalization is also consistent with
Wilson's modern interpretation of quantum field theory, where quantum field theory can be defined up to some finite energy scale no matter whether there exists a UV completion or whatever the would-be UV completion is. So it is intriguing to refine such a connection based on the recently developed Wilsonian formulation of holographic renormalization\cite{HP,FLR}.

On the other hand, to our best knowledge, the conserved current method we have used in the proof of our bulk/boundary correspondence is totally novel in the context of holography. This method further suggests another natural local correspondence between the black hole horizon and boundary by the integral curves of conserved current. Such a local correspondence seems more reasonable than the conjectured one defined by the null geodesics in the previous literature such as \cite{Bhattacharyya,CMSM,CMY}\footnote{It is noteworthy that in our holography what we are concerned with is the entropy production while in previous literature the concerned quantity is the so called entropy current whose divergence gives rise to the entropy production. There are at least two reasons for us to prefer the entropy production to the entropy current. For one thing, although the entropy current is sort of elegant object, its definition has some kind of ambiguity in nature. For another, as far as we know, the entropy current has never shown up in any kind of dynamics except its divergence, namely the entropy production.}.

We conclude with various other issues worthy of further investigation. For one thing, we have worked simply to second order perturbation so far. It is interesting to see whether the whole procedure can be performed to any higher order. For another, we have worked merely within the context of Einstein's gravity. It is worthwhile to see whether our holography can also be valid for other higher derivative gravity theories, where the entropy is given by Wald formula\cite{Wald3,IW}. We hope to address these issues elsewhere.

\begin{acknowledgments}
We thank Sijie Gao, Yi Ling, and Junbao Wu for helpful discussions. HZ  would like to thank Elias Kiritsis, Rene Meyer, Takeshi Morita, Ioannis Papadimitriou, and Tassos Petkou for their valuable discussions. He is also grateful to the organizers of Spring School  on Superstring Theory and Related Topics for their financial support and fantastic hospitality at ICTP, where the relevant discussions with participants, in particular with Hong Liu and Shiraz Minwalla  are much appreciated. This
work is partly supported by the National Natural Science Foundation
of China (Grant Nos. 11075206 and 11175245). It is also supported by European Union grants FP7-REGPOT-2008-1-CreteHEP Cosmo-228644, and PERG07-GA-2010- 268246 as well as EU program ÓThalisÓ ESF/NSRF 2007-2013.
\end{acknowledgments}

\end{document}